\documentclass[preprint,showpacs,preprintnumbers,nofootinbib,amsmath,amssymb,aps,prc,longbibliography]{revtex4-1}
\usepackage{graphicx}
\usepackage{dcolumn}
\usepackage{bm}
\begin{document}
\title{Specific shear viscosity in hot rotating systems of paired fermions}
\author{N. Quang Hung$^{1}$}
 \altaffiliation[On leave of absence from Center for Theoretical and Computational Physics, ]
 {College of Education, Hue University, Vietnam}
 \email{hung.nguyen@ttu.edu.vn}
\author{N. Dinh Dang$^{2,3}$}
  \email{dang@riken.jp}
 \affiliation{1) School of Engineering, TanTao University, TanTao University Avenue, TanDuc Ecity, Duc Hoa, Long An Province, Vietnam \\
2) Theoretical Nuclear Physics Laboratory, RIKEN Nishina Center
for Accelerator-Based Science,
2-1 Hirosawa, Wako City, 351-0198 Saitama, Japan\\
3) Institute for Nuclear Science and Technique, Hanoi, Vietnam}

\date{\today}
\begin{abstract}
The specific shear viscosity $\overline\eta$ of a classically rotating system of nucleons that interact via a monopole pairing interaction is calculated including the effects of thermal fluctuations and coupling to pair vibrations within the selfconsistent quasiparticle random-phase approximation.  It is found that $\overline\eta$ increases with angular momentum $M$ at a given temperature $T$. In medium and heavy systems,  $\overline\eta$ decreases with increasing $T$ at $T\geq$ 2 MeV and this feature is not affected much by angular momentum. But in lighter systems (with the mass number $A\leq$ 20), $\overline\eta$ increases with $T$ at a value of $M$ close to the maximal value $M_{max}$, which is defined as the limiting angular momentum for each system. The values of $\overline\eta$ obtained within the schematic model as well as for systems with realistic single-particle energies are always larger than the universal lower-bound conjecture $\hbar/(4\pi k_B)$ up to $T$=5 MeV.

\end{abstract}

\maketitle
\section{Introduction}
\label{Intro}
Viscosity is an important mechanical property of fluids, which describes their resistance to flow. The recent ultrarelativistic Au-Au and Pb-Pb collisions at the Relativistic Heavy Ion Collider (RHIC) at Brookhaven National Laboratory~\cite{RHIC} and the Large Hadron Collider (LHC) at CERN~\cite{LHC} have revealed a strongly interacting matter that behaves like a nearly perfect fluid with extremely low viscosity. This has generated a high interest in the study of viscosity in various systems, which consist of strongly interacting particles. In nuclear physics, although viscosity has been theoretically calculated and experimentally extracted in 1970s, the calculations of the specific shear viscosity $\overline\eta\equiv\eta/s$, that is the ratio of shear viscosity $\eta$ to the entropy volume density $s$, in finite nuclei as a function of temperature $T$ was reported very recently only in two papers~\cite{Auerbach,viscosity}. The results of these works, which have been carried out in two different approaches, show that the specific shear viscosity in hot nuclei at 
a temperature $T$ as high as 5 MeV is actually very close to what obtained in the strongly interacting matter discovered at RHIC and LHC and $T>$ 170 MeV. The approach in Ref. \cite{viscosity} used the Green-Kubo relation to calculate the shear viscosity $\eta$ of a finite hot nucleus directly from the width and energy of the giant dipole resonance (GDR) of this nucleus. The Green-Kubo relation expresses the shear viscosity in terms of the correlation function of the shear stress tensors. As a result, the shear viscosity can be calculated from the imaginary part of the retarded Green's function that describes the transport process.

The studies in Refs. \cite{Auerbach,viscosity} consider systems at zero angular momentum. How the shear viscosity changes in a hot rotating finite system is an interesting question. Recently the approach in Ref. \cite{viscosity} has been extended to calculate the specific shear viscosity from the parameters of the giant dipole resonance (GDR) in hot rotating nuclei~\cite{GDRJ}. Following the same line, in the present paper we would like to study the specific shear viscosity of a finite system of paired fermions, which interact via the monopole pairing force at finite angular momentum. We first consider a schematic model with doubly folded equidistant levels, then apply the formalism to several realistic nuclei. Despite being illustrated by the results of calculations within nuclei, the present formalism can be applied to any finite systems of fermions with discrete single-particle energies interacting via a monopole pairing force.

 The paper is organized as follows. The formalism is derived in Sec. \ref{formalism}. The analysis of numerical results obtained in the schematic multilevel model with pairing as well as by using realistic single-particle spectra for $^{20}$O, $^{44}$Ca, and $^{120}$Sn are discussed in Sec. \ref{results}. The paper is summarized in the last section, where conclusions are drawn.
\section{Formalism}
\label{formalism}
We consider the pairing Hamiltonian describing a spherical system, which is classically 
rotating about the symmetry $z$ axis \cite{Moretto}:
\begin{equation}
H = H_{P}-\lambda\hat{N}-\gamma\hat{M}~.  \label{H}
\end{equation}
Here $\lambda$ and $\gamma$ are the chemical potential and rotation frequency, respectively. $H_{P}$ 
describes a system of $N$ fermions, which 
interact via a monopole pairing force with the constant parameter $G$, namely
\begin{equation}
H_P=\sum_k\epsilon_k(a_{+k}^{\dagger}a_{+k}+a_{-k}^{\dagger}a_{-k}) - G\sum_{kk'}{a_k^{\dagger}
a_{-k}^{\dagger}a_{-k'}a_{k'}}~, \hspace{5mm} (k~{\rm and}~k' >0)~,\label{Hpair}
\end{equation}
where $a_{\pm k}^{\dagger}(a_{\pm k})$ are the creation (annihilation) operators of a nucleon (neutron or proton) with angular momentum $k$, projection $\pm m_k$ ($m_k >$ 0), and energy $\epsilon_k$. The particle-number operator $\hat{N}$ and total angular momentum $\hat{M}$, which coincides with its $z$ projection, are given as
\begin{equation}
\hat{N}=\sum_k(a_{+k}^{\dagger}a_{+k} + a_{-k}^{\dagger}a_{-k})~, \hspace{5mm}
\hat{M}=\sum_k m_k(a_{+k}^{\dagger}a_{+k} - a_{-k}^{\dagger}a_{-k}) ~. \label{NM}
\end{equation}
By using the Bogolyubov transformation $a_k^\dagger = u_k\alpha_k^\dagger + v_k\alpha_{-k}~$ from the particle operators, $a_k^\dagger$ and  $a_k$ to the quasiparticle ones, $\alpha_k^\dagger$ and $\alpha_k$,  
the Hamiltonian \eqref{H} is transformed into the quasiparticle one $\cal H$, whose explicit form can be found, e.g., in Refs. \cite{SCQRPA,SCQRPAJ}.

The Hamiltonian (\ref{H}) has been recently solved within the FTBCS1+SCQRPA in Refs.~\cite{SCQRPA, SCQRPAJ}.
The FTBCS1+SCQRPA includes the effect caused by the quasiparticle-number fluctuation (QNF) at finite temperature (FT), which is neglected in the standard BCS solution of the pairing problem, as well as the effect due to coupling to pairing vibration within the selfconsistent quasiparticle random-phase approximation (SCQRPA). Since the FTBCS1+SCQRPA has already been discussed thoroughly in Refs. \cite{SCQRPA,SCQRPAJ,SCQRPAJ1}, we summarize below only their main results, which are necessary for calculations in the present paper. 
\subsection{FTBCS1 equations at finite angular momentum}
\label{FTBCS1}

The FTBCS1 equations are a set of coupled equations for the pairing gap, particle number and angular momentum. The abbreviation "FTBCS1" denotes that, different from the conventional FTBCS, QNF is taken account in FTBCS1.
The equation for the level-dependent pairing gap $\Delta_{k}$ is given as
\begin{equation}
\Delta_k = \Delta + \delta\Delta_k ~, \label{Gap}
\end{equation}
with the level-independent gap $\Delta$ and the level-dependent gap $\delta\Delta_k$ defined as
\begin{equation}
\Delta=G\sum_{k'}{u_{k'}v_{k'}(1-n_{k'}^+-n_{k'}^-)}~, \hspace{5mm}
\delta\Delta_k=G\frac{\delta{\cal N}_k^2}{1-n_k^+-n_k^-}u_kv_k ~, \label{Gapeq1} 
\end{equation}
where
\begin{eqnarray}
u_k^2&=&\frac{1}{2}\left(1+\frac{\epsilon_k-Gv_k^2-\lambda}{E_k}\right)~, \hspace{5mm}
v_k^2=\frac{1}{2}\left(1-\frac{\epsilon_k-Gv_k^2-\lambda}{E_k}\right)~, \nonumber \\
E_k&=&\sqrt{(\epsilon_k-Gv_k^2-\lambda)^2+\Delta_k^2}~, \hspace{5mm}
 \label{uv}
\end{eqnarray}
and $\delta{\cal N}_k^2$ is the QNF at nonzero angular momentum
\begin{equation}
\delta{\cal N}_k^2=(\delta{\cal N}_k^+)^2+(\delta{\cal N}_k^-)^2 = n_k^+(1-n_k^+)+n_k^-(1-n_k^-) ~. \label{QNF}
\end{equation}
The equations for the particle number $N$ and total angular momentum $M$ are [See Eqs. (25) and (26) of Ref. \cite{SCQRPAJ}]: 
\begin{equation}
N = 2\sum_k\left[v_k^2(1-n_k^+ -n_k^{-}) + \frac{1}{2}(n_k^++n_k^{-}) \right], \hspace{5mm}
M = \sum_k m_k(n_k^+ - n_k^{-}) ~. \label{NM1}
\end{equation}
In these equations $n_k^{\pm}$ are quasiparticle occupation numbers at temperature $T$ and angular momentum projection $\mp m_{k}$ given by the Fermi-Dirac distribution of noninteracting fermions
\begin{equation}
n_k^{\pm}=\frac{1}{{\rm exp}[\beta(E_k\mp\gamma m_k)]+1} ~, \hspace{5mm} 
\beta=\frac{1}{T} ~. \label{nkpm}
\end{equation}
By neglecting the QNF, i.e., setting $\delta{\cal N}_k^2=0$ in Eq. \eqref{QNF}, the FTBCS1 equations become the conventional FTBCS ones.

\subsection{FTBCS1+SCQRPA}
\label{FTSCQRPA}

Coupling to pair vibrations beyond the FTBCS1 is carried out by solving the SCQRPA equations for the pair vibration generated by the phonon operators
 \begin{equation}
        {\cal Q}_{\mu}^{\dagger}=\sum_{k}\frac{{\cal X}_{k}^{\mu}\alpha_k^{\dagger}\alpha_{-k}^{\dagger}-
        {\cal Y}_{k}^{\mu}\alpha_{-k}\alpha_{k}}{\sqrt{1-n_k^{+}-n_k^{-}}}
        ~,\hspace{5mm} {\cal Q}_{\mu}=[{\cal
        Q}_{\mu}^{\dagger}]^{\dagger}~.
        \label{Q}
    \end{equation}
The quasiparticle Hamiltonian ${\cal H}$ is then represented in the effective form as \cite{SCQRPAJ}
\begin{equation}
        {\cal H}_{\rm eff}=\sum_{k}b_k^{+}{\cal N}_k^{+} + \sum_{-k}b_k^{-}{\cal N}_k^{-}
        +\sum_{k'}q_{kk'}{\cal N}_k{\cal N}_{k'}
        +\sum_{\mu}\omega_{\mu}{\cal Q}_{\mu}^{\dagger}{\cal Q}_{\mu}+\sum_{k\mu}V_{k}^{\mu} {\cal N}_{k}({\cal Q}_{\mu}^{\dagger}+{\cal Q}_{\mu})~,
        \label{Heff}
    \end{equation}
where ${\cal N}_k^{\pm} = \alpha_{\pm k}^{\dagger}\alpha_{\pm k}$~,  
${\cal N}_k={\cal N}_k^{+}+{\cal N}_k^{-}$, $\omega_\mu$ are the phonon energies (eigenvalues of the SCQRPA equations) and the vertex $V_k^{\mu}$ is given as
    \begin{equation}
        V_{k}^{\mu}=\sum_{k'}g_{k}(k')\sqrt{1-n_k^{+}-n_k^{-}}
        ({\cal X}_{k'}^{\mu}+{\cal Y}_{k'}^{\mu})~.
        \label{Vertex}
    \end{equation}
    The explicit expressions of $b_k^{\pm}$, $g_k(k')$ and $q_{kk'}$ are given in Eq. (10) of Ref. \cite{SCQRPAJ}, Eqs. (12) and (14) of Ref. \cite{SCQRPA}, respectively. The set of FTBCS1 and  SCQRPA
equations are solved selfconsistently to define the pairing gap $\Delta_k$, chemical potential $\lambda$, rotation frequency $\gamma$, phonon energies $\omega_{\mu}$, phonon amplitudes
${\cal X}_k^{\mu}$ and ${\cal Y}_k^{\mu}$ as well as the quasiparticle and phonon occupation numbers $n_k^{\pm}$ and $\nu_\mu$ at each values of temperature $T$ and angular momentum $M$.

After solving the FTBCS1+SCQRPA equations,  the Green's function for the quasiparticle propagation is found by using the effective Hamiltonian (\ref{Heff}) as
 \begin{equation}
        G_{\pm k}(E)=\frac{1}{2\pi}\frac{1}{E-E_{k}^{\pm}-M_{k}^{\pm}(E)}~,
        \label{GkE}
    \end{equation}
with the modified quasiparticle energies 
    \begin{equation}
        E_k^{\pm} = b_k^{\pm} + q_{kk}~,
        \label{Ek+-}
        \end{equation}
        and the mass operator, which has the analytic continuation into the complex energy plane as
        \begin{equation}
        M_{k}^{\pm}(\omega\pm i\varepsilon)=M_{k}^{\pm}(\omega)\mp i\gamma_{k}^{\pm}(\omega)~.
        \label{Mk}
    \end{equation}
    The real and imaginary parts of this analytic continuation define, respectively, the energy shift and damping of the quasiparticle due to coupling to the SCQRPA, namely
    \begin{equation}
        M_{k}^{\pm}(\omega)=\sum_{\mu}(V_{k}^{\mu})^{2}
        \bigg[\frac{(1-n_{k}^{\pm}+\nu_{\mu})(\omega-E_{k}^{\pm}-\omega_{\mu})}
        {(\omega-E_{k}^{\pm}-\omega_{\mu})^{2}+\varepsilon^{2}}+
        \frac{(n_{k}^{\pm}+\nu_{\mu})(\omega-E_{k}^{\pm}+\omega_{\mu})}
        {(\omega-E_{k}^{\pm}+\omega_{\mu})^{2}+\varepsilon^{2}}\bigg]~,
        \label{Momega}
    \end{equation}
    \begin{equation}
        \gamma_{k}^{\pm}(\omega)=\varepsilon
        \sum_{\mu}(V_{k}^{\mu})^{2}
        \bigg[\frac{1-n_{k}^{\pm}+\nu_{\mu}}
        {(\omega-E_{k}^{\pm}-\omega_{\mu})^{2}+\varepsilon^{2}}+
        \frac{n_{k}^{\pm}+\nu_{\mu}}
        {(\omega-E_{k}^{\pm}+\omega_{\mu})^{2}+\varepsilon^{2}}\bigg]~.
        \label{gamma}
    \end{equation}
In Eqs. (\ref{Mk}) -- (\ref{gamma}),  $\nu_\mu = \langle{\cal
Q}_{\mu}^{+}{\cal Q}_{\mu}\rangle$ is the phonon occupation number, which is found by solving a set of coupled equations (31) -- (35) in Ref. \cite{SCQRPAJ}, and $\varepsilon$ is a sufficient small parameter.

The spectral intensities
$J_{k}^{\pm}(\omega)$ are derived from the relations~\cite{Zubarev} 
\begin{equation}
J_{k}^{\pm}(\omega)[\exp(\omega/T)+1]/2=i[G_{\pm k}(\omega+i\varepsilon)-G_{\pm
k}(\omega-i\varepsilon)]/2=-{\rm Im}G_{\pm k}(\omega)~,
\label{ImG}
\end{equation}
in the form
\begin{equation}
J_{k}^{\pm}(\omega)=\frac{1}{\pi}
        \frac{\gamma_{k}^{\pm}(\omega)(e^{\omega/T}+1)^{-1}}         
        {[\omega-E_{k}^{\pm}-M_{k}^{\pm}(\omega)]^{2}+[\gamma_{k}^{\pm}(\omega)]^2}~,
        \label{Jk+-}
        \end{equation}
and, the quasiparticle occupation numbers $n_k^{\pm}$ are found as
    \begin{equation}
        n_{k}^{\pm}=\int_{-\infty}^{\infty}J_{k}^{\pm}(\omega)d\omega~.
        \label{nkcoupling}
    \end{equation}
In the limit of small quasiparticle damping
$\gamma_k^{\pm}(\omega)\rightarrow 0$, $n_k^{\pm}$ can be
approximated with the Fermi-Dirac distribution
    \begin{equation}
        n_{k}^{\pm}\simeq\frac{1}{{\rm exp}(\widetilde{E}_{k}^{\pm}/T)+1}~,
        \label{nklimit}
    \end{equation}
where $\widetilde{E}_{k}^{\pm}$ are the solutions of the equations for the
poles
of the quasiparticle Green's functions $G_{\pm k}(E)$ (\ref{GkE}), namely
    \begin{equation}
        \widetilde{E}_{k}^{\pm}-E_{k}^{\pm}-M_{k}^{\pm}(\widetilde{E}_{k}^{\pm}) = 0~.
        \label{Etilde}
    \end{equation}
A simplified version of FTBCS1+SCQRPA is called the FTBCS1+QRPA. 
In the latter, by ignoring the selfconsistency, the FTBCS1 is solved first. The obtained $u_k$, $v_k$,
 FTBCS1 quasiparticle energies $E_k$ and rotation frequency $\gamma$ are then used to solve the QRPA equations to determine the phonon energies $\omega_\mu$ as well as the amplitudes ${\cal X}^{\mu}_k$ and ${\cal Y}^{\mu}_k$. The quasiparticle occupation numbers are approximated with the Fermi-Dirac distribution (\ref{nklimit}), where $E_k\mp\gamma m_k$ are used in place of $\widetilde{E}_{k}^{\pm}$ (See Eq. (24) of Ref. \cite{SCQRPAJ}), and the phonon occupation numbers $\nu_\mu$ are approximated with the Bose-Einstein distribution 
 $\nu_\mu=[\exp(\omega_\mu/T)-1]^{-1}$. 
\subsection{Specific shear viscosity}
 \label{viscosity}       
 As has been mentioned in the Introduction, the specific shear viscosity is defined as 
a function of $T$ and $M$ as
\begin{equation}
\overline{\eta}(T,M)=\frac{\eta(T,M)}{s(T,M)}~,\hspace{5mm} s(T,M) =\frac{\rho}{A}S(T,M)~,
\label{r}
\end{equation}
where $\eta(T,M)$ is the shear viscosity and $s(T,M)$ is the entropy density at finite $T$ and $M$, $\rho$ is the nuclear density, $A$ is the number of nucleons, and $S(T,M)$ is the entropy of the system. From Eq. (3) of Ref. \cite{viscosity}, we have
\begin{equation}
\eta(T,M) = \eta_0\frac{\lim_{\omega\rightarrow 0}{\rm Im}G(\omega,T,M)}{\lim_{\omega\rightarrow 0}{\rm Im}G(\omega,0,0)}~.
\label{eta1}
\end{equation}
A normalization  is carried out in Eq. (\ref{eta1}) so that in the limit of $T=$ 0 and $M=$ 0, the value of $\eta(T,M)$ is equal to the parameter $\eta_0$, whose value can be extracted from experimental systematics.

By using Eq. (\ref{ImG}), we obtain from Eq. (\ref{eta1})
\begin{equation}
\eta(T,M) = 
\eta_0\frac{J(0,T,M)}{J(0,0,0)}~.
\label{eta2}
\end{equation}
In Eq. (\ref{eta2}) we introduce the shorthand notation
\begin{equation}
J(0,T,M)=\sum_kJ_k(0,T,M)/\Omega~,\hspace{5mm}  J_k(0,T,M) =J_k^{+}(0,T,M) + J_k^{-}(0,T,M)~,
\label{J0TM}
\end{equation}
where $\Omega$ is the sum of all single-particle levels.\footnote{For the spherical case: $J(0,T,M)=\sum_j\Omega_jJ_j(0,T,M)/\sum_j\Omega_j$ with $\Omega_j=j+1/2$.} 
Replacing the right-hand side of Eq. (\ref{eta2}) with its explicit expression obtained by using Eq. (\ref{Jk+-}) with the mass operator (\ref{Momega}) and quasiparticle damping (\ref{gamma}), we
obtain the final expression for $\eta(T,M)$ in the following form
 \begin{equation}
   \eta(T,M) = \eta_0\frac{{J}^{+}(T,M)+{J}^{-}(T,M)}{2{J}(0,0)}~,
   \label{eta3}
   \end{equation}
   with the explicit expressions for ${J}^{\pm}(T,M)$ and ${J}(0,0)$ given as
   \begin{equation}
   {J}^{\pm}(T,M) =\sum_k\frac{\varepsilon\sum_{\mu}(V_{k}^{\mu})^{2}
        \bigg[\frac{1-n_{k}^{\pm}+\nu_{\mu}}
        {(E_{k}^{\pm}+\omega_{\mu})^{2}+\varepsilon^2}+
        \frac{n_{k}^{\pm}+\nu_{\mu}}
        {({E}_{k}^{\pm}-\omega_{\mu})^{2}+\varepsilon^2}\bigg]}
        {(\widetilde{E}^{\pm}_k)^2+\varepsilon^2\bigg\{\sum_{\mu}(V_{k}^{\mu})^{2}
        \bigg[\frac{1-n_{k}^{\pm}+\nu_{\mu}}
        {(E_{k}^{\pm}+\omega_{\mu})^2+\varepsilon^2}+
        \frac{n_{k}^{\pm}+\nu_{\mu}}
        {(E_{k}^{\pm}-\omega_{\mu})^2+\varepsilon^2}\bigg]\bigg\}^2}~,
        \end{equation}
        \begin{equation}
        { J}(0,0)=\sum_k\frac{\varepsilon
        \sum_{\mu}\frac{(V_{k}^{\mu})^2}
        {\big[b_k(0)+q_{kk}(0)+\omega_{\mu}(0)\big]^2 +\varepsilon^2}}
        {[\widetilde{E}_k^{\pm}(0)]^2+\varepsilon^2
        \bigg\{
        \sum_{\mu}\frac{(V_{k}^{\mu})^{2}}
        {\big[b_{k}(0)+q_{kk}(0)+\omega_{\mu}(0)\big]^2+\varepsilon^2}\bigg\}^2}~,
 \end{equation}
 where  $b_k(0)$, $q_{kk}(0)$, $\widetilde{E}_k^{\pm}(0)$, and $\omega_\mu (0)$ are, respectively, the values of $b_k$, $q_{kk}$, solutions of Eq. (\ref{Etilde}) and SCQRPA eigenvalues, determined at $T=$ 0 and $M=$ 0.   
 
 The entropy $S(T,M)$ is calculated as the sum of the quasiparticle and phonon entropies as
 \begin{equation}
 S(T,M) = S_{\alpha}(T,M) + S_{\cal Q}(T,M)~,
 \label{S}
 \end{equation}
 where the quasiparticle entropy $S_{\alpha}(T,M)$ is given as
 \begin{equation}
 S_{\alpha}(T,M)=-\sum_k\big[n_k^+\ln n_k^++(1-n_k^+)\ln(1-n_k^+)+
 n_k^-\ln n_k^-+(1-n_k^-)\ln(1-n_k^-)\big]~,
 \label{Sq}
 \end{equation}
 whereas the phonon entropy is defined as the boson one, namely
 \begin{equation}
 S_{\cal Q}(T,M)=\sum_\mu\big[(1+\nu_\mu)\ln(1+\nu_\mu)-\nu_\mu\ln \nu_\mu\big]~.
 \label{Sp}
 \end{equation}

Within the FTBCS (FTBCS1), the mass operator $M_k^{\pm}(\omega)$, quasiparticle damping $\gamma_k^{\pm}(\omega)$, phonon occupation numbers $\nu_{\mu}$, and phonon energies $\omega_{\mu}$ are zero. Therefore, the expressions for $J^{\pm}(T,M)$ and $J(0,0)$ are given as
\begin{equation}
J^{\pm}(T,M)=\sum_k{\frac{1}{(E_k^{\pm})^2+\varepsilon^2}}~, \hspace{5mm}
J^{\pm}(0,0)=\sum_k{\frac{1}{[E_k^{\pm}(0)]^2+\varepsilon^2}}~, \label{etaFTBCS}
\end{equation}
where $E_k^{\pm}=E_k\pm \gamma m_k$ with $E_k$ being the quasiparticle energies obtained within the FTBCS (FTBCS1).
\section{Analysis of numerical results}
\label{results}
The numerical calculations are carried out for the schematic model as well as realistic nuclei. 
The schematic model consists of $\Omega$ doubly folded equidistant levels, which interact via a pairing force with parameter $G$.
Before switching on the pairing interaction, $\Omega/2$ lowest levels are occupied by $N=\Omega$ particles with 2 particles on each levels with spin projections $\pm m_k = \pm 1/2, \pm 3/2,...,\pm\Omega-1/2$, whereas $\Omega/2$ upper levels are empty.
The distance between levels is chosen equal to 1 MeV \cite{SCQRPAJ}.
As for the realistic nuclei, $^{20}$O, $^{44}$Ca, and $^{120}$Sn are considered. The single-particle spectra for these nuclei are obtained within the axially deformed Woods-Saxon potential including the spin-orbit and Coulomb interactions~\cite{WS}. All the bound (negative energy) single-particle states are taken into account in the calculations.
The pairing interaction parameter $G$ is adjusted so that the pairing gap at $T=0$ reproduces the experimental value obtained from the odd-even mass difference. 
These nuclei have the proton closed shell, so there is only neutron pairing gap. Therefore $G_N$ is chosen equal to 1.04 MeV, 0.53 MeV, and 0.14 MeV, which yields $\Delta_N(0) \approx $ 3.0 MeV, 2.0 MeV, and 1.42 MeV for $^{20}$O, $^{44}$Ca, 
and $^{120}$Sn, respectively. Since the pairing gap (\ref{Gap}) is level-dependent, in the analysis we plot the level-weighted gap $\bar{\Delta}$, which is defined as
$\bar{\Delta}=\sum_k\Delta_k/\Omega$ with $\Omega$ being the total number of levels. 

In the calculations of the shear viscosity $\eta$ we adopt the value of nuclear density $\rho=$ 0.16 fm$^{-3}$. A value $\varepsilon=$ 0.5 MeV in Eqs. (\ref{Momega}) and (\ref{gamma}) is used in all calculations to ensure the smooth behavior of the obtained $\eta(T,M)$. Regarding the value $\eta(0,0)$, in principle, a superfluid has zero viscosity at $T=$ 0 and $M=$ 0. At $T\neq$ 0 (or $M\neq$ 0), the pairing gap decreases with increasing $T$ (or $M$) whereas the admixture of normal mode appears because of the QNF. As a result the shear viscosity $\eta(T,M)$ becomes finite at $T\neq$ 0 (or M$\neq$ 0). As for the realistic nuclei under consideration, they consist of proton closed shells and neutron opened shells. Therefore the shear viscosity is finite even at $T=$ 0 because of the protons in the normal phase. Meanwhile, according to the lower bound  $\eta/s\geq\hbar/(4\pi k_B)$ conjectured in Ref. \cite{KSS}, the shear viscosity $\eta(T,M)$ cannot be 0 at $T\neq$ 0 even at very small $T$ because $s$ is always positive at any $T\neq$ 0. In other words, 
there are no perfect fluids, that is fluids having no shear viscosity, at $T\neq$ 0.
Since in the present paper we are interested in the behavior of $\eta(T,M)$ as a function of $T$ and $M$, we adopt  $\eta(0,0)\equiv\eta_0=$ 1.0$\times$ 10$^{-23}$ Mev s fm$^{-3}$ in the calculations for realistic nuclei as well as for the schematic model.  This value is obtained by fitting the width of giant resonances at $T=$ 0 (See Ref. \cite{viscosity} for the detail discussions on the parameter $\eta_0$).

    \begin{figure}
       \includegraphics[width=13.5cm]{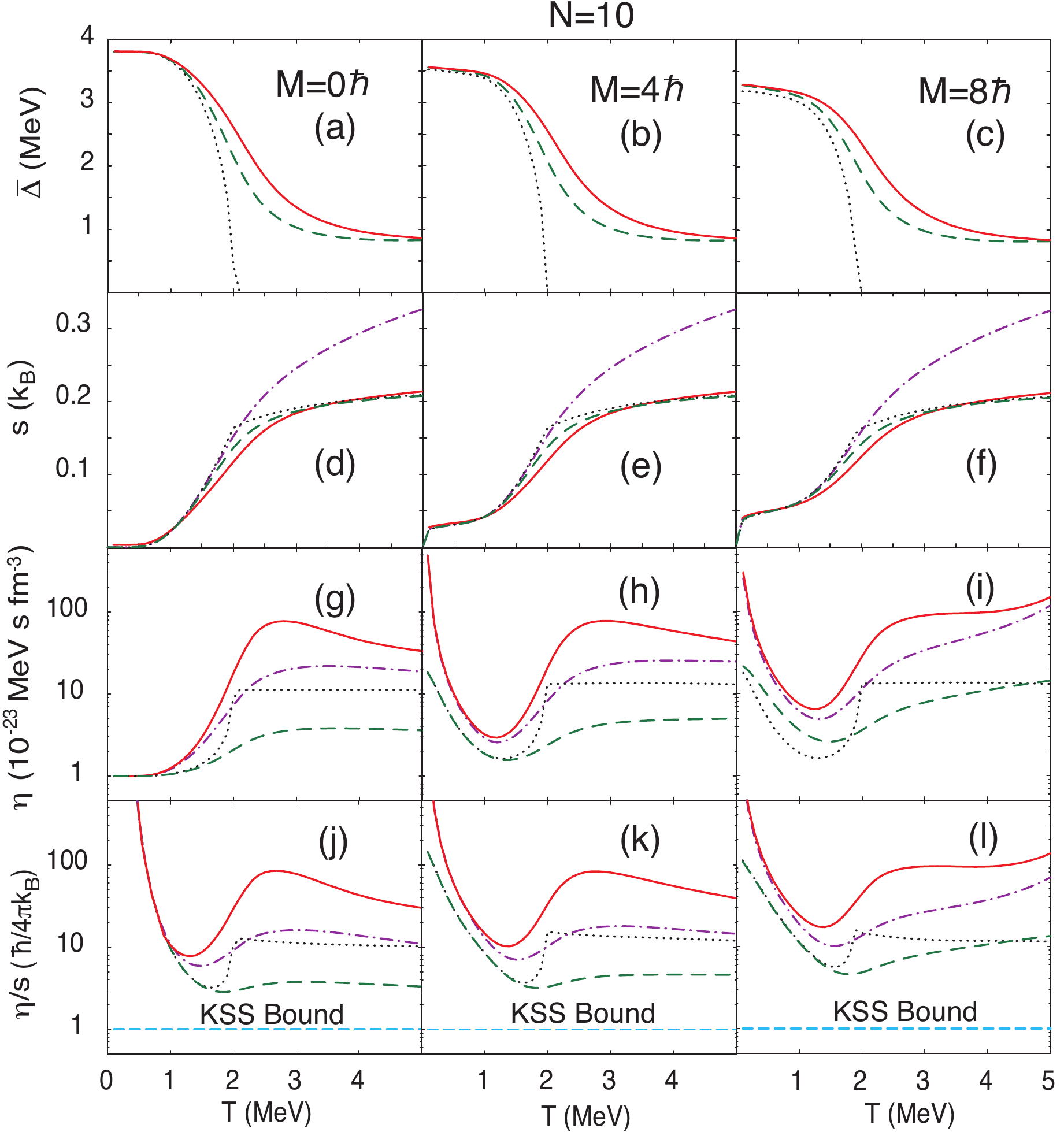}
        \caption{(Color online)  Level-weight pairing gap $\bar{\Delta}$, density entropy $s(T,M)$, shear viscosity $\eta(T,M)$, and specific shear viscosity $\overline{\eta}(T,M)\equiv\eta(T,M)/s(T,M)$ obtained for the schematic model with $N=$ 10 within the FTBCS (dotted line), FTBCS1 (dashed line),
FTBCS1+QRPA (dash-dotted line), and FTBCS1+SCQRPA (solid line) at different values of angular momentum $M$. The horizontal dashed lines in  (j) - (l) denotes the lower bound conjecture  $\hbar/(4\pi k_B)$ proposed in Ref. \cite{KSS}.
        \label{N10}}
    \end{figure}
    \begin{figure}
       \includegraphics[width=13.5cm]{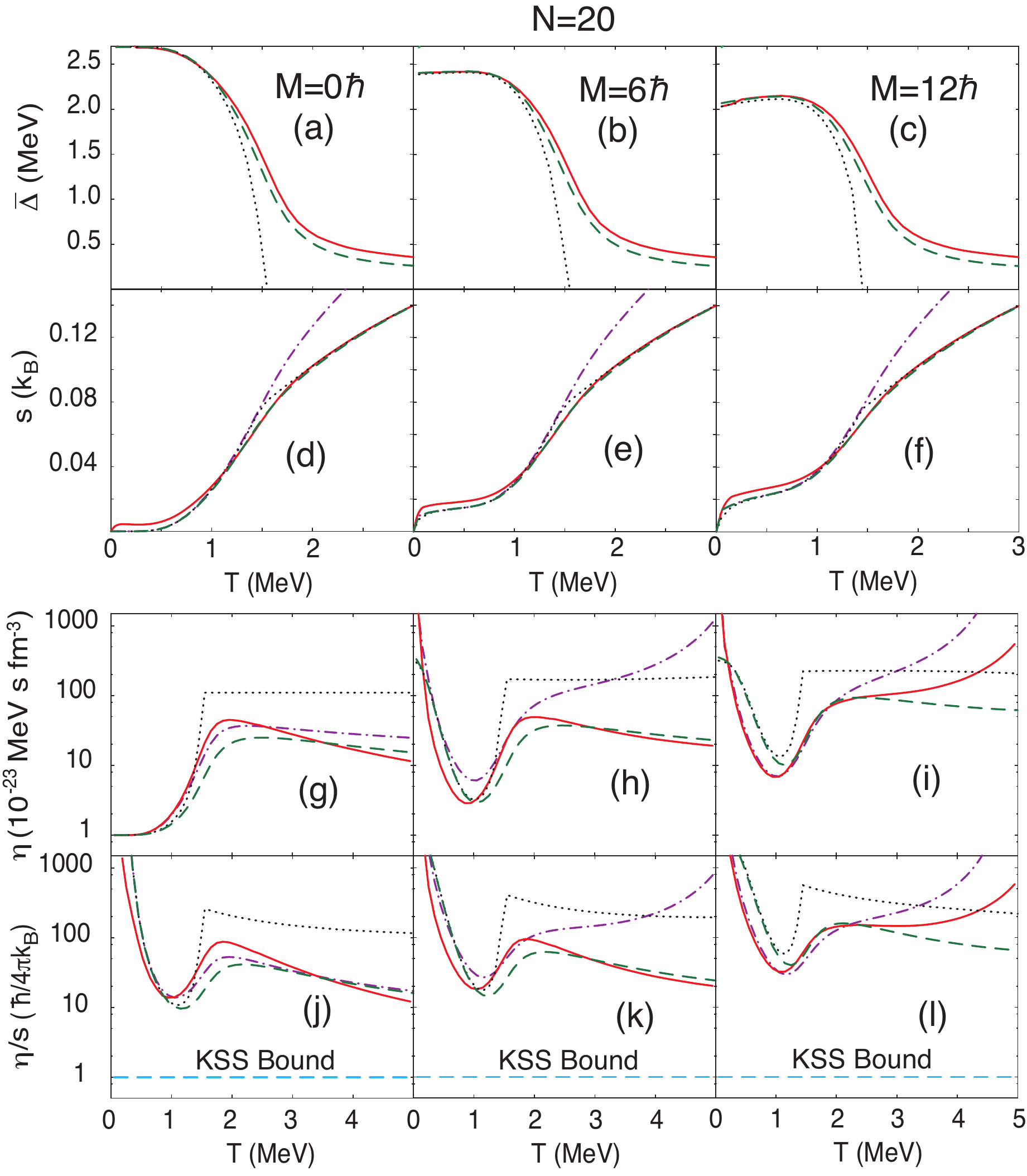}
        \caption{(Color online)  The same as in Fig. \ref{N10} but for N=20.
        \label{N20}}
    \end{figure}
    \begin{figure}
       \includegraphics[width=14cm]{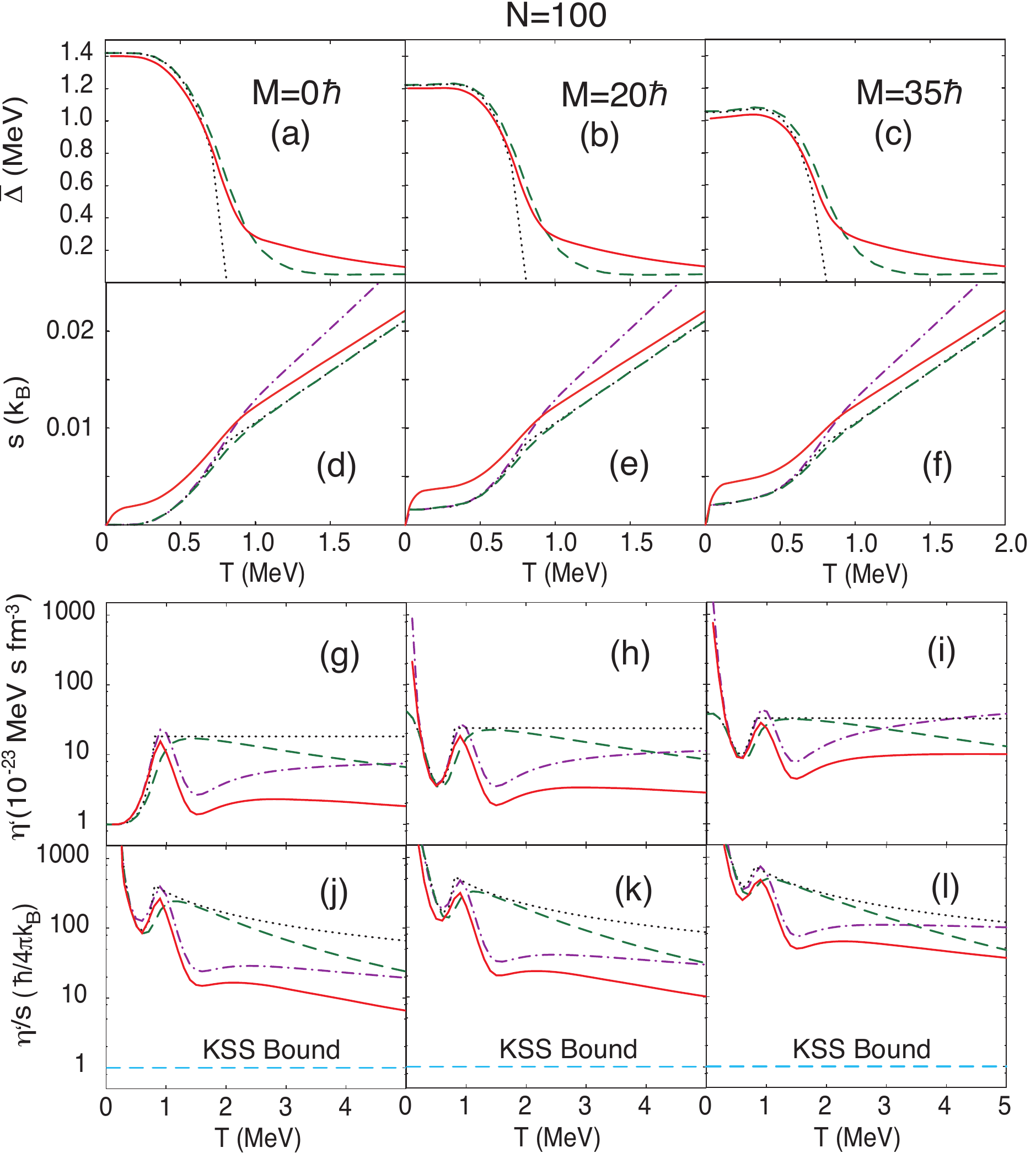}
        \caption{(Color online)  The same as in Fig. \ref{N10} but for N=100.
        \label{N100}}
    \end{figure}
Shown in Figs. \ref{N10} -- \ref{N100} are the level-weight pairing gap $\bar{\Delta}$, density entropy $s(T,M)$, shear viscosity $\eta(T,M)$, and specific shear viscosity $\overline{\eta}(T,M)\equiv\eta(T,M)/s(T,M)$ obtained as functions of $T$ at 3 values of  angular momentum $M$ for the schematic model with $N=$ 10, 20 and 100, respectively. As has been discussed in Refs. \cite{SCQRPAJ} and references therein, a distinguished feature of the FTBCS1 is that the gap $\bar{\Delta}$ never collapses as the FTBCS gap at the critical temperature $T_c$ of the phase transition from the superfluid phase to the normal one, but, because of QNF, it  decreases monotonically with increasing $T$, and has a tail up to $T=$ 5 MeV [Panels (a) -- (c)]. With increasing $N$, the tail becomes depleted and eventually vanishes at $N\rightarrow\infty$. The gap decreases with increasing $M$ as well known within the conventional FTBCS. The corrections due to coupling to SCQRPA slightly increases the pairing gap in the temperature region around the BCS phase transition, but this effect quickly decreases with increasing $N$ to become negligible at $N>$ 100.

The entropy densities predicted by the FTBCS, FTBCS1, and FTBCS1+SCQRPA are close to each other, although the SCQRPA corrections slightly decrease the entropy at $T<$ 3 MeV for N = 10, and increase it at all $T$ for $N\geq$ 20 [Panels (d) -- (f)]. Meanwhile, the FTBCS1+QRPA entropy density differs strongly with these results at high $T$ because the Bose-Einstein distribution, which approximates the phonon occupation number $\nu_\mu$ within the QRPA, differs significantly from $\nu_\mu$ obtained by solving the selfconsistent equations (31) -- (35) in Ref. \cite{SCQRPAJ} employed within the SCQRPA (See e.g. Fig. (5) of Ref. \cite{SCQRPA}). By comparing the panels (d) -- (f) in these figures, one can see that angular momentum has negligible effect on the entropy density.

Within the FTBCS, the shear viscosity $\eta$ increases with $T$ from its value $\eta_0$ at $T=$ 0 to reach at $T>T_c$ a value larger than $\eta_0$ by one to two orders of magnitude, demonstrating that the normal fluid is highly viscous and its viscosity remains essentially constant in the normal phase [Panels (g) -- (i)]. QNF within the FTBCS1 significantly decreases $\eta$ at $T>T_c$, whereas the effects due coupling to the QRPA and SCQRPA increase and decrease $\eta$ for $N\leq 20$ and $N>20$, respectively. The largest difference between the FTBCS1 and FTBCS1+SCQRPA is almost 2 orders of magnitude at $T\sim$ 2.7 MeV. Increasing the angular momentum leads to an overall increase of $\eta$, especially at low $T$. In other words, rotation makes a systems of fermions that interact via monopole pairing force becomes more viscous. As the particle number $N$ increases, $\eta$ obtained within the FTBCS1+SCQRPA (QRPA) decreases to approach the prediction by the FTBCS1 [See Fig. \ref{N20} (g) -- \ref{N20} (i)], which increases with $N$. In large systems. e.g. $N=$100, $\eta$ obtained within the FTBCS1+SCQRPA (QRPA) becomes even smaller than that predicted by the FTBCS1 [See Fig. \ref{N100}]. The combine effect of temperature and angular momentum leads to a local minimum in the temperature dependence of $\eta$ at $T$ slightly below $T_c$ and local maximum at $T$ slightly above $T_c$. With increasing $N$ this feature transforms into a local maximum at $T\simeq T_c$ with a slight local minimum at each side. 

The overall trend of the specific shear viscosity $\overline\eta$ is a decrease with increasing $T$ in the region $T<T_c$. At $T$ around $T_c$ a local maximum is seen, which is the direct consequence of the local maximum in the temperature dependence of $\eta$ discussed above. At $T\gg T_c$ the specific shear viscosity $\overline{\eta}$ obtained with the FTBCS1+SCQRPA for large $N$ ($\geq$ 100) also decreases with increasing $T$. 

    \begin{figure}
       \includegraphics[width=13.5cm]{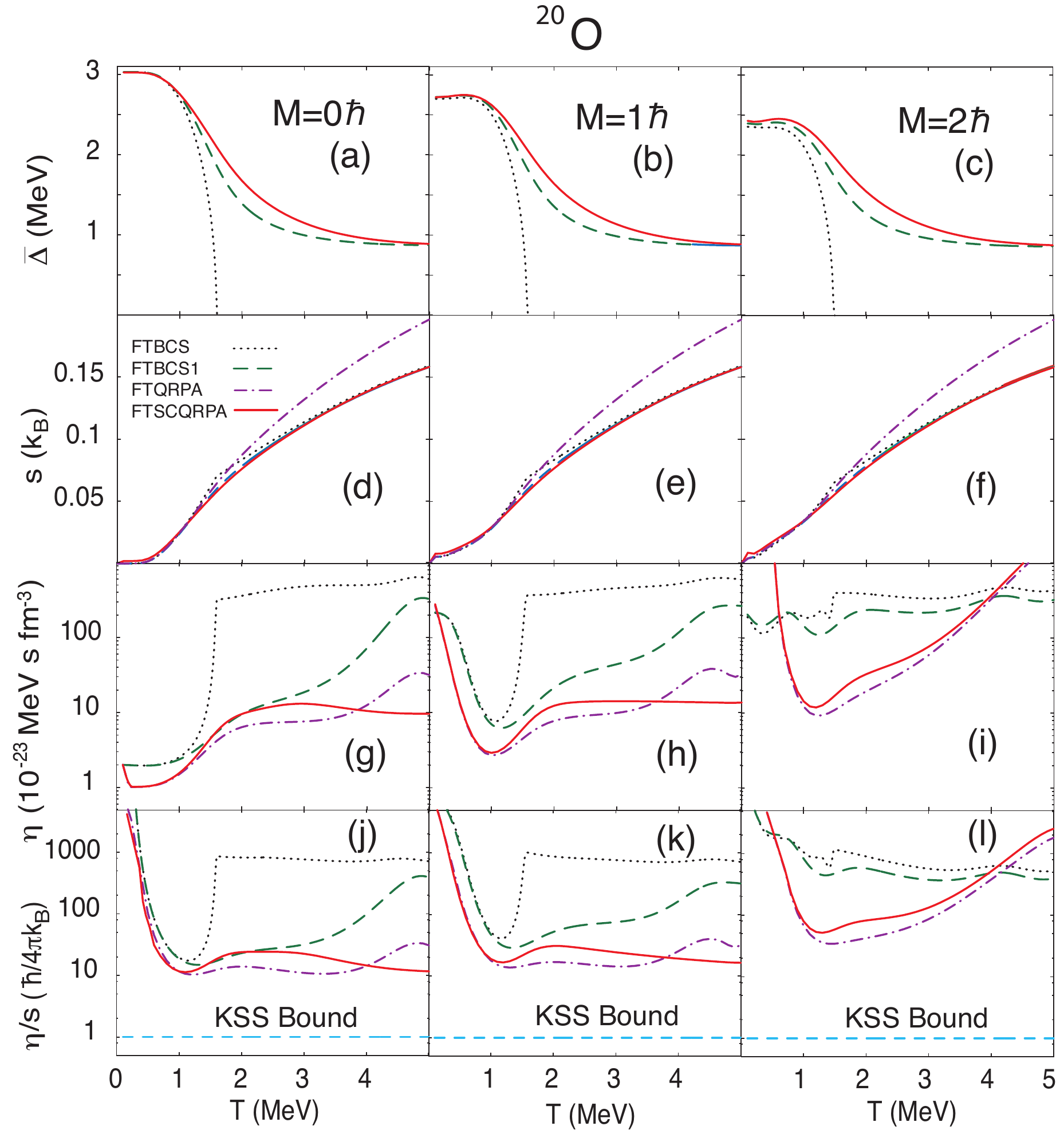}
        \caption{(Color online)  The same as in Fig. 1 but for $^{20}$O. 
        \label{O}}
    \end{figure}
    \begin{figure}
       \includegraphics[width=13.5cm]{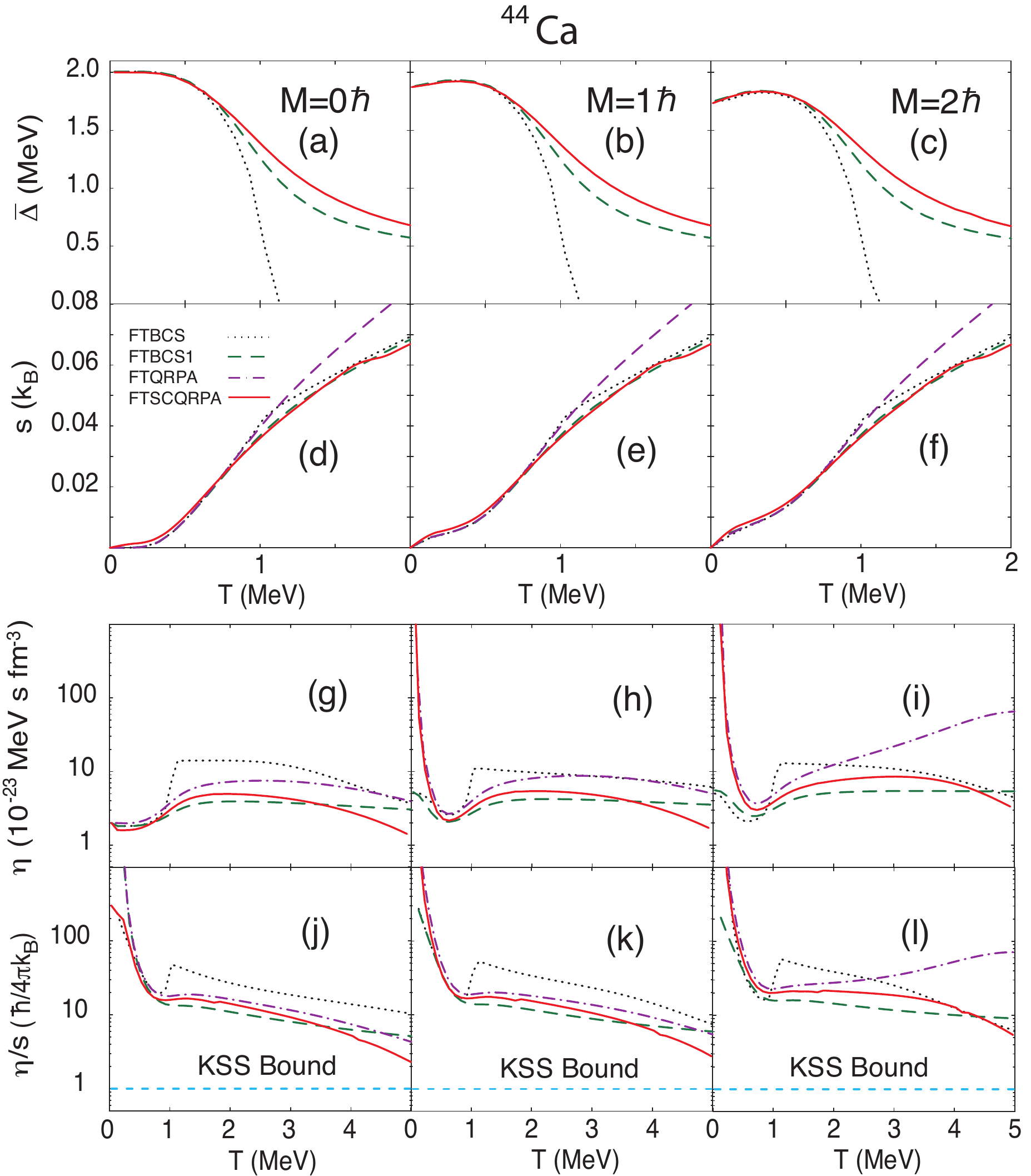}
        \caption{(Color online)  The same as in Fig. 1 but for $^{44}$Ca.
        \label{Ca}}
    \end{figure}
    \begin{figure}
       \includegraphics[width=13.5cm]{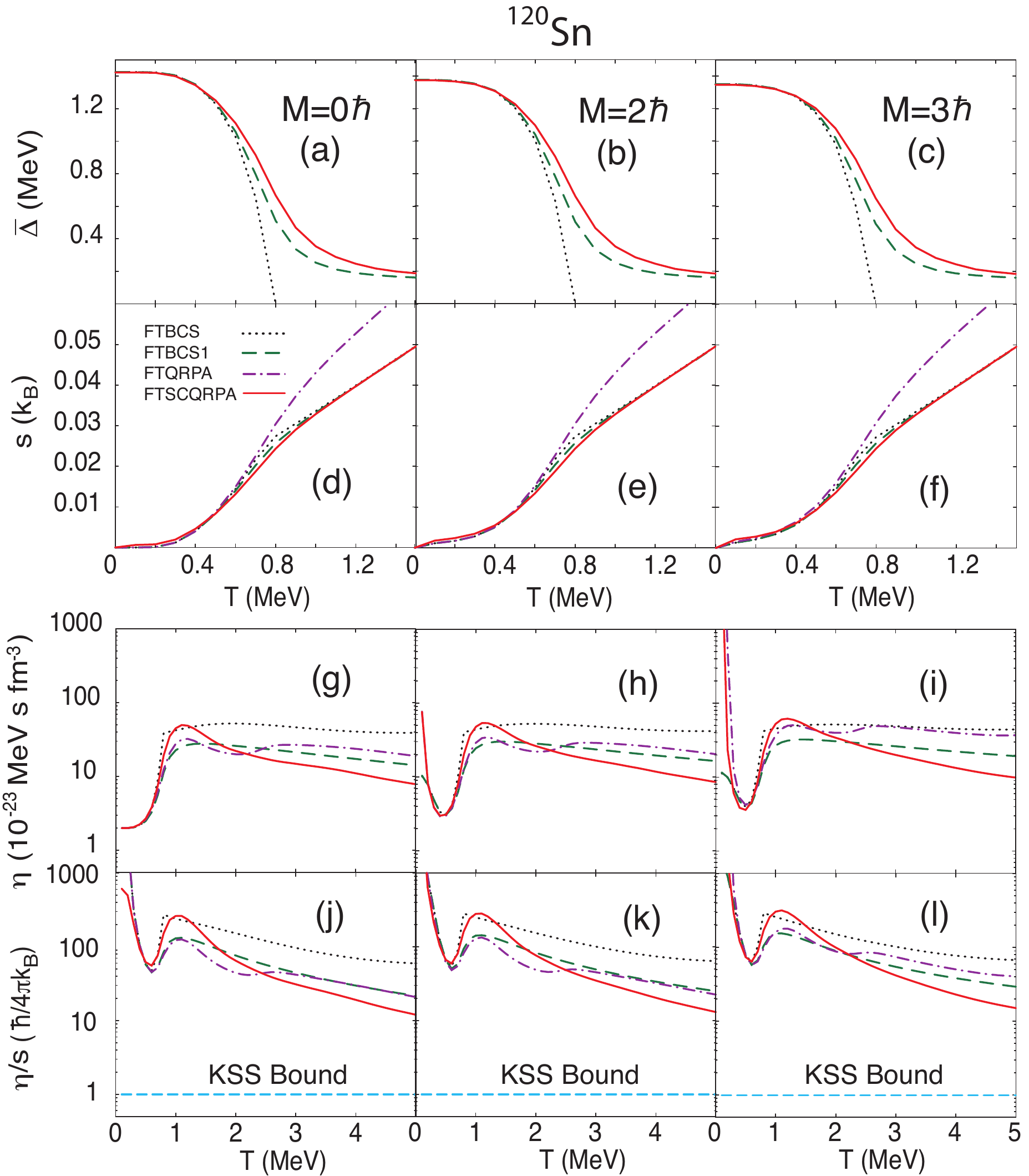}
        \caption{(Color online)  The same as in Fig. 1 but for $^{120}$Sn.
        \label{Sn}}
    \end{figure}
The qualitative features obtained for the shear viscosity with the schematic model as a function of $T$ at various $M$ also hold for realistic nuclei, namely $^{20}$O, $^{44}$Ca, and $^{120}$Sn, as shown in Figs. \ref{O} -- \ref{Sn}. In a light system such as $^{20}$O, QNF and coupling to SCQRPA (QRPA) cause dramatic effects, which significantly reduce $\eta(T,M)$ as compared to the prediction within the FTBCS (FTBCS1) in the temperature region $T_c\leq T\leq$ 3.5 MeV. For $^{44}$Ca and $^{120}$Sn, the shear viscosity $\eta$ always decreases  with increasing $T$ at all $M$ except for the prediction by the FTBCS1+QRPA at $M=$ 2$\hbar$ for $^{44}$Ca because of the boson occupation numbers discussed above. These figures show that, at $T=$ 5 MeV and $M=$0, the specific shear viscosity $\overline\eta$ in spherical systems of paired fermions with realistic single-particle energies ranges from between twice to ten times larger than its value at $T=$ 0 and $M=$ 0.  Classical rotation strongly increases the specific shear viscosity in the light system such as $^{20}$O at hight $T$, whereas in the heavier systems such as $^{44}$Ca and $^{120}$Sn such increase is much weaker. The change in the temperature dependence of $\eta$ in $^{20}$O at high $T$ from decreasing as $T$ increases at low $M$ to increasing with $T$ at large $M$ shows that  rotation of a light system at high $T$ drives it from the behavior of a liquid to that of a gas. Meanwhile, medium and heavy systems behave as liquids because the predictions by the FTBCS1+SCQRPA show that their shear viscosity always decreases as $T$ increases at high $T$ and any $M$. In general, it is clear to see that the specific shear viscosity obtained within the schematic model as well as realistic nuclei considered here is always higher than the lower bound conjecture (KSS limit) proposed in Ref. \cite{KSS} up to $T$=5 MeV.

    \begin{figure}
       \includegraphics[width=16cm]{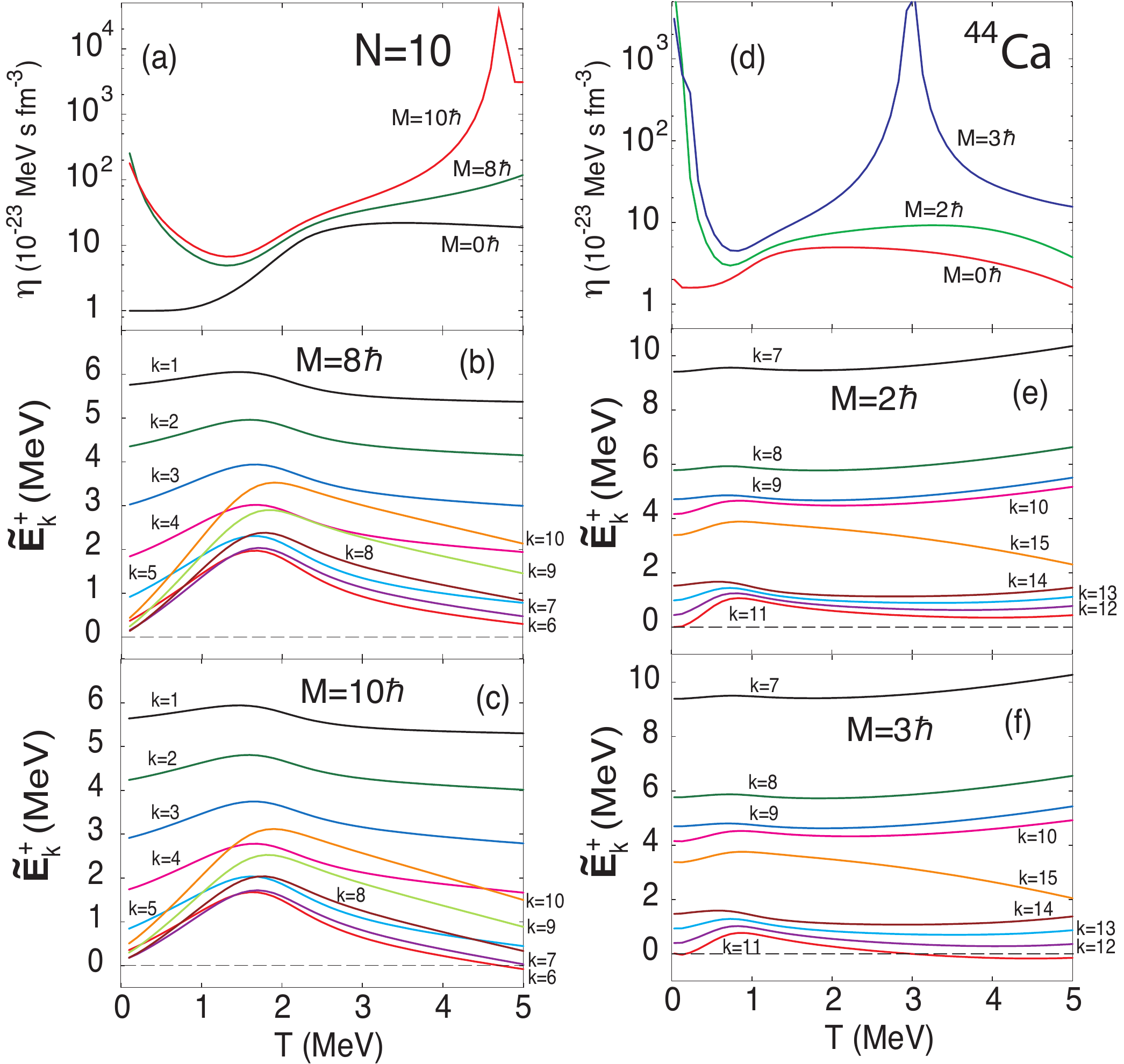}
        \caption{(Color online)  Shear viscosity $\eta(T,M)$ and quasiparticle energies $\widetilde{E}_{k}^{\pm}(T,M)$ (\ref{Etilde}) for several levels within the schematic model with $N=$ 10 [(a) -- (d)] and for neutrons in $^{44}$Ca [(d) -- (f)].
        \label{Etest}}
    \end{figure}

For small $N$,  FTBCS1+SCQRPA (QRPA) predicts an increase $\overline{\eta}$ with $T$ at the largest value of $M$, namely 8 and 12 $\hbar$ for $N=$ 10 and 20,  respectively. This largest value should be smaller than the value $M_{\rm max}$, at which at least one solution of Eq. (\ref{Etilde}) becomes 0 at a certain $T$, and turns negative at higher $T$. This is demonstrated in Fig. \ref{Etest}, where $\eta$ and $\widetilde{E}_{k}^{+}$ for several levels within the schematic model for $N=$ 10 and for neutrons in $^{44}$Ca are plotted against $T$ at three values of $M$. For $N=$ 10, at $M_{\rm max}=$ 10$\hbar$ one can see that $\eta$ has a singularity at  $T\sim$ 4.7 MeV [Fig. \ref{Etest} (a)], where $\widetilde{E}_{k}^{+}$ crosses 0 [Fig. \ref{Etest} (c)]. For $^{44}$Ca, a similar feature takes place at $T=$ 3 MeV and $M_{\max}=$ 3$\hbar$ [Figs. \ref{Etest} (d) and \ref{Etest} (f)]. The quasiparticle energies $\widetilde{E}_{k}^{-}$  remain positive at all $T$ and $M$ because $E_k^{-}$ in Eq. (\ref{Ek+-}) are always positive. The limiting values $M_{\rm max}$ of the angular momentum, determined for $N=$ 20, 100, $^{20}$O, and $^{120}$Sn, are 13$\hbar$, 43$\hbar$, 3$\hbar$ and 4$\hbar$, respectively. That's is the reason why only the results obtained for $M<M_{\rm max}$ are shown in Figs. \ref{N10}--\ref{Sn}. 

It is important to keep in mind that the results and analyses in the present paper are obtained by using the Hamiltonian (\ref{H}), which describes essentially the motion of quasiparticles in the quasiparticle mean field, modified by rotation and coupling to vibration of monopole quasiparticle pairs. In general, they are not valid for nuclear systems whose residual interactions consist of nonzero multipolarities such as dipole, quadrupole, octupole, etc, since the latter correspond to the Green's functions different from Eq. \eqref{GkE}, which lead to different behaviors of the shear viscosity $\eta(T,M)$. As a matter of fact, the specific shear viscosity $\overline{\eta}$ extracted in Ref. \cite{viscosity} based on the parameters of GDR in hot rotating nuclei shows a tendency of decreasing with increasing $M$ at high $T$. However, the present work provides a theoretical formalism, which is based on the spectral intensities or response functions to derive and calculate the shear viscosity for not only nuclear but also any finite hot rotating systems.
\section{Conclusions}
In the present paper, the Green-Kubo relation is used to calculate the specific shear viscosity from the retarded Green's function that describes the propagation of quasiparticles within the quasiparticle mean field of a classically rotating system of nucleons that interact via a monopole interaction. Thermal fluctuations such as the QNF are included within the FTBCS1, whereas coupling to monopole pair vibrations is taken into account within the SCQRPA.

The general feature of the specific viscosity $\overline\eta$ of this system can be summarized as follows.  At a given temperature $T$, $\overline\eta$ increases with the angular momentum $M$, that is a rotating system of paired fermions is more viscous. In medium and heavy systems,  $\overline\eta$ decreases with increasing $T$ at $T\geq$ 2 MeV and this feature is not affected much by angular momentum. However, in light systems, it increases with $T$ at the values of angular momentum $M$ close to $M_{\rm max}$, at which the quasiparticle energy $\widetilde{E}_{k}^{+}$ (\ref{Etilde}) crosses 0 at a certain $T$. 
 At $T<$ 2 MeV, local minima and/or local maximum appear because of the significant change in the curvature
of the temperature dependence of the thermal pairing gap. Thermal fluctuations and coupling to the quasiparticle pair vibrations within the SCQRPA significantly increases $\overline\eta$ for small N systems with $N\le 10$, whereas  $\overline\eta$ decreases for large $N>10$ systems. All the results of $\overline\eta$ obtained within the schematic model as well as realistic nuclei are always larger than the universal lower bound of the specific shear viscosity up to $T$=5 MeV.

Although the present paper considers the pairing model only, it lays the way of calculating the shear viscosity in not only nuclear but also any finite hot rotating Fermi systems based on the spectral intensities or response functions, and therefore opens the possibility to study the evolution of the shear viscosity as a function of temperature and angular momentum in these systems.

\acknowledgments
The numerical calculations were carried out using the FORTRAN IMSL
Library by Visual Numerics on the RIKEN Integrated Cluster of Clusters (RICC) system. NQH acknowledges the support by the National Foundation for Science and Technology Development(NAFOSTED) of Vietnam through Grant No. 103.04-2010.02. 


\begin{thebibliography}{99}
  \bibitem{RHIC}K. Adcox {\it et al.} (PHENIX Collaboration), Nucl. 
    Phys. A {\bf 757}, 184 (2005); B.B. Back {\it et al.}, Ibid. 
    {\bf 757}, 28 (2005); J. Arsene {\it et al.} (BRAHMS Collaboration),
    Ibid. {\bf 757}, 1 (2005); J. Adams {\it et al.} (STAR
    Collaboration), Ibid. {\bf 757}, 102 (2005).
    \bibitem{LHC}K. Arnold {\it et al.} (ALICE Collaboration), 
    Phys. Rev. Lett. {\bf 105}, 252302 (2010); G. Aad {\it et al.}
    (ATLAS Collaboration), Ibid. {\bf 105}, 252303 (2010).
 \bibitem{Auerbach}N. Auerbach and S. Shlomo, Phys. Rev. Lett. {\bf
    103}, 172501 (2009).
 \bibitem{viscosity}N. Dinh Dang, Phys. Rev. C {\bf 84}, 034309 (2011).   
 \bibitem{GDRJ}N. Dinh Dang, Phys. Rev. C {\bf 85}, 064323 (2012).
 \bibitem{Moretto}T. Kammuri, Prog. Theor. Phys. {\bf 31}, 595 (1964); 
 L. G. Moretto, Nucl. Phys. A \textbf{185}, 145 (1972).
  \bibitem{SCQRPA}N. Dinh Dang and N. Quang Hung, Phys. Rev. C {\bf 77}, 064315 (2008).
 \bibitem{SCQRPAJ}N.Q. Hung and N.D. Dang, Phys. Rev. C {\bf 78}, 064315 (2008).
 \bibitem{SCQRPAJ1}N.Q. Hung and N.D. Dang, Phys. Rev. C {\bf 84}, 054324 (2011).
 \bibitem{WS}S. Cwiok {\it et al.}, Comput. Phys. Commun. {\bf 46}, 379 (1987).
  \bibitem{Zubarev}D.N. Zubarev, Sov. Uspekhi {\bf 3}, 320 (1960)
    [Usp. Fiz. Nauk. {\bf 71}, 71 (1960)].
    \bibitem{KSS}P.K. Kovtun, D.T. Son, and A.O. Starinets, Phys. Rev. Lett.
    {\bf 94}, 111601 (2005).     
\end{thebibliography}
\end{document}